\documentclass[twocolumn]{revtex4}
\usepackage{amssymb}
\usepackage{amsmath}
\usepackage{mathptmx}
\usepackage{graphicx}
\usepackage{epsfig}
\usepackage[colorlinks]{hyperref}

\begin{document}

\title{\bf Single-gap superconductivity in Mo$_8$Ga$_{41}$}

\author{ M.\,Marcin,$^{1}$ J.\,Ka\v cmar\v c\'ik,$^1$ Z.\,Pribulov\'a,$^1$  M.\,Kop\v c\' ik,$^{1}$ P.\,Szab\'o,$^1$ O. \v Sofranko,$^1$ T. Samuely,$^1$ V. Va\v no,$^2$   C.\,Marcenat,$^3$  V.Yu.\,Verchenko,$^{4,5}$ A.V.\,Shevelkov$^4$ and P.\,Samuely$^1$}

\affiliation{$^1$ Centre of Low Temperature Physics, Institute of Experimental Physics SAS,  and P. J. \v Saf\'arik University, 040 01 Ko\v sice, Slovakia\\
$^2$ Technical University, Letn\'a 9, SK-04001 Ko\v sice, Slovakia\\
$^3$ SPSMS, UMR-E9001, CEA-INAC/UJF-Grenoble 1, 17 Rue des martyrs, 38054 Grenoble, France\\
$^4$  Department of Chemistry, Lomonosov Moscow State University, 119991 Moscow, Russia\\
$^5$ National Institute of Chemical Physics and Biophysics, 12618 Tallinn, Estonia}

\date{\today}

\begin{abstract}
In this paper, potential two-gap superconductivity in Mo$_8$Ga$_{41}$ is addressed in detail by means of thermodynamic and spectroscopic measurements. Combination of highly sensitive ac-calorimetry and scanning tunneling spectroscopy (STS), as bulk and surface sensitive probes, utilized on the same piece of crystal reveals that there is only one intrinsic gap in the system featuring strong electron-phonon coupling. Traces of multiple superconducting phases seen by STS and also in the heat capacity measured in high magnetic fields on a high-quality and seemingly single-phase crystal might mimic the multigap superconductivity of Mo$_8$Ga$_{41}$ suggested recently in several studies. 
\end{abstract}
\maketitle

\section*{Introduction}

Two-gap superconductivity is a compelling phenomenon as it comprises new riches of condensed matter physics, including e.g. new mechanism for spontaneous symmetry breaking in systems of superconducting vortices \cite{Milosevic}, or existence of fractional vortices \cite{Babaev}. The hunt for its representatives continues since the experimental justification of two energy scales in MgB$_2$ \cite{MgB2} in 2001.  

Two-gap superconductors are characterized by the existence of two distinct energy gaps that reside on separated parts of the Fermi surface interconnected by some interband scattering. This scattering results in closing of both energy gaps at the same critical temperature $T_c$. Techniques like scanning tunneling spectroscopy (STS) are able to show two gaps directly but other measurements sensitive to quasiparticle density of states (DOS) can be usefull, too. For example, temperature dependence of the heat capacity \cite{bouquet} and of the upper critical magnetic field $H_{c2}$ \cite{Lyard} can be used to consistently model two gaps in the system. Unfortunately, a presence of additional phase(s) may sometimes reveal similar behavior mimicking a true multi-gap case. Two-gap superconductivity was proposed for example in Cu$_x$TiSe$_2$ based on the muon spin rotation experiments \cite{Zaberchik} 
 and in $\beta$-Bi$_2$Pd from the heat capacity and the positive curvature of the upper critical magnetic field \cite{Imai} but it has been disproved when a combination of highly sensitive techniques was employed in Ref.\cite{kacmarcik1},\cite{kacmarcik2}. Thus, before any definitive conclusion, a combination of techniques capable to address several aspects of the phenomenon  revealing consistent picture should be employed.

Recently, it was suggested that Mo$_8$Ga$_{41}$ features two-gap superconductivity \cite{Verchenko2, Neha, Sirohi} that vanishes upon the V for Mo substitution \cite{Verchenko2}. The material is a member of endohedral gallium cluster compounds.  In the structure of Mo$_8$Ga$_{41}$, each Mo atom is placed inside a cage of 10 Ga atoms forming endohedral clusters that share all their vertices. This architecture resembles that of perovskite oxides, among which various important superconductors, including high-$T_c$ oxides, can be found. Recently, it was also noted that superconductivity and structural stability probably compete in this family of galium-based superconductors \cite{Xie}. Therefore, unconventional features in Mo$_8$Ga$_{41}$ might be anticipated. There are several members of this family that are known to be superconducting. Besides Mo$_8$Ga$_{41}$ with $T_c$\,$\sim$ 9.8\,K \cite{Bezinge}, superconductivity was also reported  in Mo$_6$Ga$_{31}$ with $T_c$\,$\sim$ 8\,K \cite{Fischer}.

Superconducting properties of Mo$_8$Ga$_{41}$ were studied by transport and thermodynamic measurements in Ref. \cite{Verchenko1}. Heat capacity and magnetic susceptibility were measured on a collection of single crystals glued together in order to obtain reasonable signal. For transport measurements, polycrystalline samples were used. Indications of the strong-coupling superconductivity in the system were found. In the subsequent study by means of muon spin rotation/relaxation spectroscopy, it was suggested  \cite{Verchenko2} that two superconducting energy gaps exist in Mo$_8$Ga$_{41}$. Later on, the reports on possible existence of two energy gaps in Mo$_8$Ga$_{41}$ followed from the critical current \cite{Neha} and STS measurements \cite{Sirohi}.

Here, we present a comprehensive study of superconductivity in individual tiny  single crystals of Mo$_8$Ga$_{41}$ by means of bulk and surface sensitive methods. ac-calorimetry is employed to study fine structure of the heat capacity anomaly at the superconducting transition by sweeping temperature or magnetic field, while scanning tunneling microscopy (STM) and STS are used to directly probe the superconducting gaps. Our analysis based on the heat capacity measurements shows that the system is clearly a single-gap superconductor, however, traces of other minor superconducting phases may mimic the multigap behavior in the surface-sensitive techniques as evidenced by the STS data.

\section*{Experiment}

Single crystals of Mo$_8$Ga$_{41}$ were synthesized using the flux growth method. Details of synthesis can be found in the previous report \cite{Verchenko1}. The obtained crystals were characterized by a combination of electron probe x-ray microanalysis and single-crystal x-ray diffraction, and no deviations from the Mo$_8$Ga$_{41}$ composition and crystal structure were found. The residual-resistance-ratio of $\text{RRR}=15.4$ found in the standard electrical transport measurements indicates good quality of the crystals.

Thermodynamic properties were measured by the ac-calorimetry using a light emitting diode as a contact-less source of heating power \cite{ac-calorimetry}. Individual crystals were glued on a chromel-constantan thermocouple, which served both as a sample holder and a thermometer to detect oscillations of the sample temperature. Correction of the Cernox thermometer and the thermocouple in magnetic field were carefully inspected and accounted for during the data treatment. In order to subtract the addenda from the total heat capacity, the empty thermocouple was measured in zero magnetic field and in fields up to 10 T. In the experiment, the heat was supplied to the sample at a frequency of several Hertz. Measurements were performed down to 600 mK in $^3$He cryostat in 8 T horizontal, and 10 T vertical magnets. The magnetic field was applied both perpendicular and parallel to the flat sample facet sticked to the thermocouple. As we did not find any significant difference in a position of the superconducting transition for the two magnetic field orientations (the difference was less than 1$\%$),  we consider system to be isotropic.

Superconducting properties of the sample surface were probed by STM. Prior to the experiment to avoid any contamination, the surface was treated either ex situ by polishing on an  Al$_2$O$_3$  plate or in situ by mild Ar$^+$ sputtering in the Specs UHV STM system. Subsequent STM and STS measurements performed at the base temperature of 2 K showed no noticeable difference between the two surface treatment methods. Therefore, further STM and STS experiments were performed on the ex situ treated samples employing our homemade STM system immersed in a Janis SSV $^3$He cryostat allowing measurements down to 400 mK in magnetic fields up to 8 T. A gold STM tip was used for measurements. The spectroscopy measurements were performed by obtaining current-voltage ($I-V$) characteristics, then numerically differentiating the $I-V$ curves to acquire the tunneling conductance spectra  $G(V)$\,=\,d$I(V)$/d$V$ and normalizing those to the normal-state conductance $G_N$. Tunneling spectra were fitted by the tunneling conductivity model for the normal metal-superconductor (N-I-S) tunneling junction\cite{STM}, with the thermally smeared BCS density of states of the superconducting electrode. Two-gap spectra were tested by fitting the data by the convolution of two BCS conductance spectra, assuming two different energy gaps $\Delta$($T$)  with complementary weights ($w_1$ and $w_2$\,=\,1 - $w_1$, respectively). Spectral conductance maps were measured at $T$\,=\,450 mK in zero magnetic field using the Current Imaging Tunneling Spectroscopy technique \cite{Hamers} with the $128\times128$ spatial resolution for a given surface area and using a bias voltage range of $\pm$\,8\,mV.

\section*{Results and Discussion}

\begin{figure*}[t]
\begin{center}
\includegraphics[width=\linewidth]{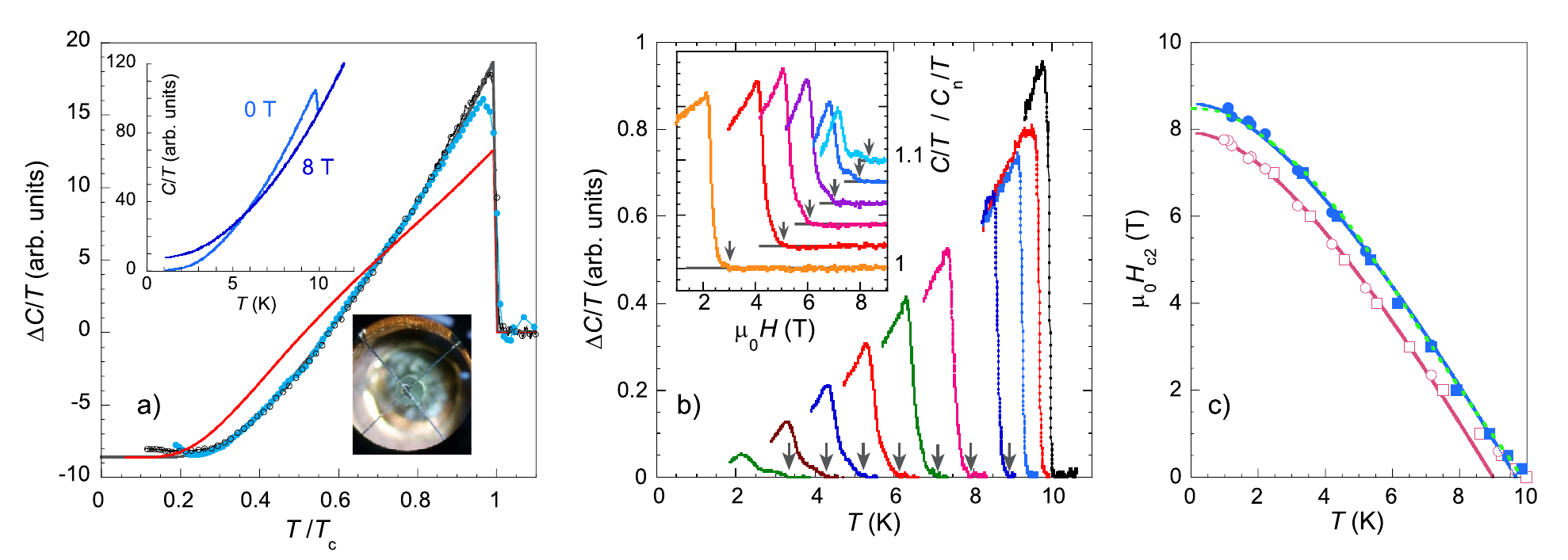}
\caption{Heat capacity of an individual single crystal of Mo$_8$Ga$_{41}$: a) Upper inset: heat capacity of the sample measured in 0 T and 8 T magnetic fields. Main panel: field-dependent part of the electronic heat capacity from the present study (empty black symbols), and from Ref.\cite{Verchenko1} (filled blue symbols), the grey line is the single-gap  $\alpha$ model curve with 2$\Delta$/k$T_c$\,=\,4.4, the red line is the two-gap  $\alpha$ model curve as described in the text. Lower inset: The sample mounted on a thermocouple with the optical fibre in the background.  b) Heat capacity after normal-state contribution subtraction, arrows point to the onset of superconductivity. Temperature sweeps measured in 7, 6, 5, 4, 3, 2, 1, 0.5, 0.2, and 0 T magnetic fields are shown from left to right. Inset: field sweeps of the heat capacity normalized to the normal-state contribution measured at 7.2, 5.2, 4.2, 3.2, 2.2, 1.7 K from left to right, curves are shifted on Y-axis for clarity. c) Upper critical field determined at the mid-point of the anomaly (empty symbols) and at the onset of the transition (filled symbols), from the temperature sweeps (squares) and field sweeps (circles). Solid lines are theoretical curves from the WHH model, the green dashed line is a theoretical curve for a strong-coupling superconductor taken from Ref.\cite{Carbotte}.}
\end{center}
\end{figure*}

Figure 1 summarizes results obtained from the heat capacity measurements. The upper panel of Fig.1a shows the total heat capacity divided by temperature $C/T$ of an individual piece of single crystal of Mo$_8$Ga$_{41}$ measured in 0 T and 8 T magnetic fields. In zero magnetic field, the anomaly at the superconducting transition is clearly visible at $T_c\simeq10$\,K. The anomaly is sharp, indicating good quality of the sample. In the 8 T measurement, no clear anomaly is present, however, further analysis of the data showed that superconductivity is not fully suppressed in overall temperature range by this magnetic field, additional small contribution is still present at low temperatures. 
For illustration, the lower inset of Fig.1a shows a snap-shot of the experimental arrangement. The diameter of the fibre is 1 mm and the sample dimensions are approximately 550 x 250 $\mu$m$^2$ with the thickness of 150 $\mu$m. 

The main panel of Fig.1a shows a plot of a field-dependent part of the electronic heat capacity from this study (empty black symbols), calculated as ${\Delta C}$/$T$ = $C$(0T)/$T$ - $C$(8T)/$T$. Subtracting the 8 T data removes lattice contribution from the total heat capacity and leaves only the field-dependent electronic part. The main panel includes also the data from the previous report measured on a polycrystalline-like sample \cite{Verchenko1} calculated in the same way (filled blue symbols).  Overlap of the two sets of data is very good except at the transition, where the superconducting anomaly of single crystal is much sharper, and in the limited range at  low temperatures. Note that not all of the data points from our present measurement (black symbols) are shown in the figure, only one point out of 50 is displayed for clarity. The solid grey line is a theoretical curve according to the $\alpha$-model \cite{alpha-model} corresponding to a single-gap superconductor with the coupling ratio of 2$\Delta$/k$T_c$\,=\,4.4. The theoretical line follows the experimental data in very good agreement, except the low-temperature region below 3\,K, where the additional contribution to the heat capacity measured in 8 T is present as mentioned above. The value of 2$\Delta$/k$T_c$\,=\,4.4 exceeds the weak coupling limit of the BCS theory, indicating strong electron-phonon coupling in the superconducting state. For comparison, we also show the solid red line that corresponds to the two-gap $\alpha$-model fit with $\Delta_1$ = 1.63 meV, $\Delta_2$ = 1.1 meV, and the weight of the larger gap of 0.8. This set of parameters was, among others, proposed by Sirohi et al. \cite{Sirohi} in order to describe tunnelling spectra of Mo$_8$Ga$_{41}$ showing two-gap features. We selected this specific combination of energy gap values and their respective weight as they result in the heat capacity behavior as close as possible to the observed one.Nevertheless, the two-gap model curve is clearly not consistent with the experimental data. Any other combination of the energy gap values proposed by Sirohi et al. leads to even more pronounced disagreement.

In Figure 1b, evolution of the heat capacity with both temperature and magnetic field is shown. Main panel of the figure depicts superconducting anomaly while sweeping the temperature at  fixed magnetic fields. At relatively low fields, the anomaly remains sharp, but when the field is increased the anomaly broadens and finally it splits in two in high magnetic fields. Arrows inserted in the figure highlight the onset of the transition. Inset of Fig.1b shows several normalized heat capacity measurements while sweeping the magnetic field at fixed temperatures, the curves are shifted on Y-axis for clarity. Again, the arrows point to the onset of superconducting transition. Note that at low temperatures (two upper curves) the anomaly reveals two distinct jumps. This is an important result indicating that even in the single crystal of Mo$_8$Ga$_{41}$ separate superconducting phases coexist. However, since the anomaly in zero magnetic field is very sharp, these superconducting phases most probably have identical critical temperatures, and they differ only slightly in their upper critical magnetic fields.

Using the temperature- and field-dependent heat capacity data, the upper critical field $\mu_0H_{c2}$ as a function of temperature was constructed (Figure 1c). Empty symbols refer to the mid-point of the anomaly, filled symbols correspond to the onset of superconducting transition marked by arrows in Fig.1b; squares are determined from the temperature sweeps and circles from the field sweeps. Solid lines are predictions according to the Werthamer, Helfand and Hohenberg (WHH) model \cite{WHH} in the absence of paramagnetic and spin-orbit contributions ($\alpha=0$, $\Delta_{so}=0$) rescaled by different factors to match the low-temperature saturation of the $\mu_0$$H_{c2}$ temperature dependence. The $\mu_0$$H_{c2}$($T$)  dependence displayed by empty symbols (at the mid-point of the transition) reveals a pronounced positive curvature and deviates from the WHH theoretical curve above 7\,K. The corresponding red WHH curve yields $T_c$\,=\,9\,K, which is significantly lower than the value observed in the heat capacity measurements in zero magnetic field. The positive curvature close to $T_c$ of the $\mu_0$$H_{c2}$ temperature dependence is usual consequence of the interplay between two gaps \cite{Lyard} and  might, at first glance, suggest two-gap behavior also in Mo$_8$Ga$_{41}$. However, the splitting of the superconducting transition observed in the heat capacity measurements suggests that several superconducting phases with different upper critical magnetic fields coexist in the sample rather than two distinct energy gaps. Indeed, if $\mu_0$$H_{c2}$ is determined at the onset of the transition (filled blue symbols in Fig.1c), including both transitions, the theoretical WHH curve follows the experimental data in good agreement. The fact that the temperature dependence of $\mu_{0}H_{c2}$ depends on a definition (midpoint, or onset  of the superconducting transition) indicates the presence of several superconducting phases in the same crystal. Here, we consider the onset of the transition to be a better criterion for the determination of $\mu_0$$H_{c2}$. Small positive curvature of $\mu_0$$H_{c2}$($T$) close to $T_c$ present even for this criterion might be either connected to the inhomogeneities, or be a consequence of the strong electron-phonon coupling. The latter is supported by the theoretical curve for a strong-coupling superconductivity taken from Ref.\cite{Carbotte} (green dashed line in Fig.1c).

\begin{figure*}[t]
\begin{center}
\includegraphics[width=\linewidth]{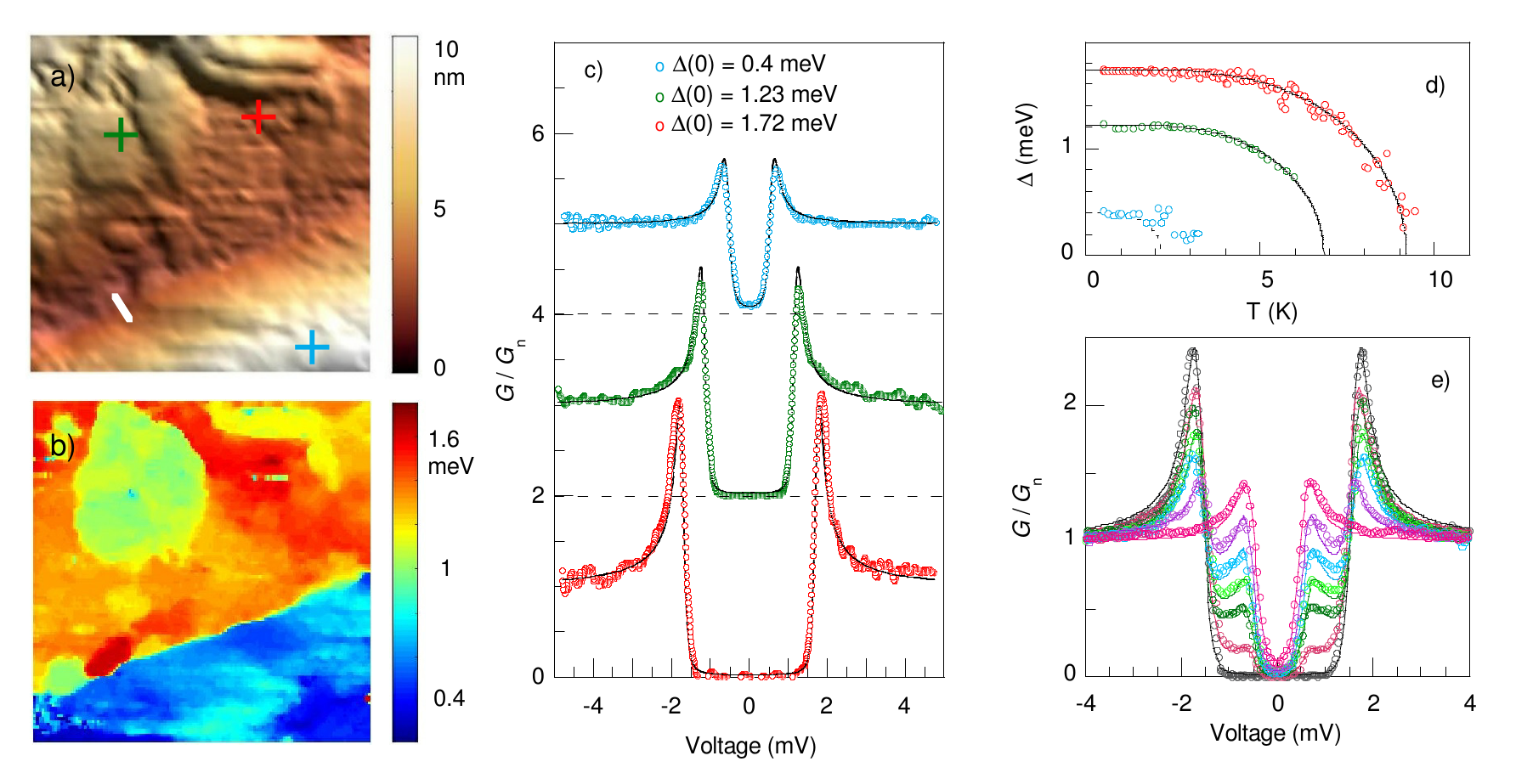}
\caption{Tunneling microscopy and spectroscopy results:  a) Surface topography  of a $200\times200$ nm$^2$ surface area at 450 mK;  b) Superconducting gap map of the area;  c) Tunneling conductance spectra at 3 points of the scanned area marked by the crosses in  Fig.2a in matching colors, the curves are shifted for clarity. The dashed lines correspond to zero conductance level. The lines are theoretical fits; d) Gap temperature dependences (symbols) measured at the positions marked by the crosses in Fig.2a and BCS fits (lines);  e) Tunneling spectra (symbols) measured along the white line in Fig.2a between two well defined phases. Lines are theoretical fits.}
\label{fig:fig2}
\end{center}
\end{figure*}

Using STM we recorded multiple surface topographs and gap maps on different parts of the sample surface covering different areas (including areas as large as $700\times700$\,nm$^2$). Surface topography scans reveal inhomogeneous surface morphology consisting of a wide variety of differently shaped and sized regions, usually protruding 10-20 nm from the surface of the sample. This is accompanied by a wide spatial distribution of the energy gap $\Delta$(0) as evidenced by tunneling spectra fits, suggesting the presence of several different surface phases. Figure 2 presents the results of topography measurements together with a spectral map measurement selected to illustrate this picture. The surface topography of a $200\times200$\,nm$^2$ area in Fig.2a shows multiple regions, which become clearly distinguished on the gap map in Fig.2b. The values of the superconducting gap $\Delta$(0) range from 0.3 meV in the protruding area to 1.75 meV in the deepest parts of the valley in the middle of the scan. Figure 2c shows individual spectra of the three main distinct areas of the scan, measured  at 450 mK after the scan was completed, with their positions marked by the crosses in Fig.2a in the matching colors.  The lines are theoretical fits giving the gap values of $\Delta$(0)\,=\,1.72,\,1.23,\, and 0.4\,meV from the bottom to the top. The temperature dependences of the energy gaps determined from three sets of spectra measured at those positions are depicted in Fig.2d by symbols, the color code is kept the same. The lines represent a standard BCS behavior. For the two larger gaps, experimental data follow the theoretical curve in good agreement with no trace of tailing effect, yielding the values of $T_c$\,=\,6.85, and 9.2\,K, which are lower than the bulk $T_c$\,=\,9.9\,K. Together with the gap values, it leads to the strong superconducting coupling ratio of 2$\Delta$/k$T_c$\,=\,4.25 $\pm$ 0.1. The different $T_c$s corresponding to different energy gaps and their almost unchanged ratio 2$\Delta$/k$T_c$ favors surface inhomogeneities scenario. If two intrinsic gaps would exist in the system, they should both close at the same $T_c$. Our measurements thus suggest that on the sample surface there are local areas with suppressed superconductivity. These areas are thicker than the coherence length of the sample ($\xi  \sim$ 6.2 nm), otherwise the proximity effect would lead to persistence of superconductivity in the junction up to the bulk $T_c$. The smallest energy gap in Fig.2c closes at $\sim$\,3\,K, which is significantly low, yet higher than what would be expected for the strong coupling. The dashed line visualizes BCS curve with 2$\Delta$/k$T_c$\,=\,4.25. Superconducting energy gap survives up to higher temperatures, probably due to some proximity effect. 

In some specific cases also two-gap spectra with two pairs of coherence peaks were observed. They were found at the boundaries between the areas with different dominant gap values as demonstrated in Fig.2e. The tunneling spectra measured along the white line in Fig.2a exhibit a convolution of two separate spectra with different gap width (1.6\,meV and 0.6\,meV) as expected for the simultaneous tunneling to two neighboring phases. The relative contributions of the larger and the smaller gaps gradually shift, while moving along the line, starting from the larger single-gap spectrum on one side (black symbols in Fig.2e), through two-gap spectra with progressivelly increasing weight of the smaller gap, ending up with the smaller single-gap spectrum (pink symbols in Fig.2e) on the other side of the line.

Even in a text-book example of the two-gap superconductivity, in MgB$_2$, two-gap spectra are not revealed at any occasion. Both energy gaps are observed only for tunneling parallel to the $ab$ plane, while in the $c$ direction, only the small energy gap was present in the spectra. This is due to different character of the Fermi surface sheets related to the two energy gaps - while $\pi$ band with the small gap is three dimensional, the large gap resides on a $\sigma$ band that is quasi two dimensional in MgB$_2$. Recently, Mo$_8$Ga$_{41}$ Fermi surface calculations were performed and presented in Ref.\cite{Sirohi} together with a report of two energy gaps existence in the system. The authors find that Fermi surface consits of several distinct sheets. Some of them are strongly anisotropic, while some other are three-dimensional. They argue that the small gap resides on a 3D sheet. This would, similarly to MgB$_2$ case, mean that the small gap  should be visible regardless the direction of the tunneling current, while the large one  could be less resolved for some directions of the current. This is actually in contradiction with our findings. In many cases we observe single-gap spectrum with only the large gap (as well as we see those for only the small gap), see e.g. Fig.2c bottom curve. If Mo$_8$Ga$_{41}$ was truly two-gap superconductor with a small gap residing on three-dimensional Fermi surface sheet, such spectrum with a large gap only could not be observed.

Moreover, in the spectra in Fig.2e, a single small gap is observed at the protrusion and a single large one at the valley, at the two extremes of the scanned line, i.e. at the same direction of the tunneling current. In between, at the slope with the current vector gradually declining to the plane of the crystal, we see two-gap spectra with a smooth transition between spectrum of small and large gaps. Such observation makes  the scenario of  MgB$_2$ where the direction of the current enables to scan different sheets of the Fermi surface in the $k$-space not applicable here. Thus, the observed two-gap spectra are not corresponding to intrinsic two-gap superconductivity, they rather reflect a convolution of two different single-gap contributions mixed in one junction as coming from neighboring areas.

From chemistry we can also take some arguments in favor of existence of several phases close to  Mo$_8$Ga$_{41}$. As was mentioned above, there is also Mo$_6$Ga$_{31}$ compound superconducting below $T_c=8$\,K \cite{Fischer}. Actually, Mo$_8$Ga$_{41}$ and Mo$_6$Ga$_{31}$  can be combined into the Mo$_n$Ga$_{5n+1}$ family of superconductors. The $n=4$ member, Mo$_4$Ga$_{21}$, was not synthesized individually. However, it can be stabilized if gallium is partially replaced by a chalcogen. Indeed, Mo$_4$Ga$_{21-y}$S$_y$, Mo$_4$Ga$_{21-y}$Se$_y$, and Mo$_4$Ga$_{21-y}$Te$_y$ compounds have been synthesized, and they show superconducting properties below $T_c\sim5$\,K in zero magnetic field \cite{mogas}. Mo$_n$Ga$_{5n+1}$ compounds exhibit a clear structural relationship. Their structures are built by MoGa$_{10}$ polyhedra and Ga$_{13}$ cuboctahedra, which are centered by unique Ga atoms. Mo$_4$Ga$_{21-y}$Ch$_y$ (Ch = S, Se, Te), Mo$_6$Ga$_{31}$, and Mo$_8$Ga$_{41}$ are individual compounds, which are superconductors below $\sim5$, 8, and 10\,K, respectively, representing different energy scales of the superconducting gaps. Their crystal structures are close to each other with the only slight difference lying in the way how MoGa$_{10}$ polyhedra are packed and organized. Note also that these phases have very close compositions, which are MoGa$_{5\frac{1}{4}}$, MoGa$_{5\frac{1}{6}}$, and MoGa$_{5\frac{1}{8}}$. Therefore, we assume that a single crystal of Mo$_8$Ga$_{41}$ may contain  the surface domains, where the packing of MoGa$_{10}$ polyhedra is slightly different resembling those in the Mo$_n$Ga$_{5n+1}$ series for $n=4$, 6, and 8. The formation of such superconducting domains may be responsible for the appearance of distinct surface regions with different superconducting energy gaps.

\section*{Conclusions}

In summary, we have performed a detailed study of superconductivity in Mo$_8$Ga$_{41}$ with $T_c$\,=\,9.9 K by means of bulk and surface sensitive techniques, i.e. ac-calorimetry and scanning tunneling microscopy/spectroscopy applied to the same single crystal. 
The heat capacity data is consistent with the single-gap $\alpha$ model with the coupling ratio of 2$\Delta$/k$T_c$ $\sim$ 4.4 and clearly excludes an existence of the second energy gap in the system. 
The heat capacity anomaly at  superconducting transition splits in two in high magnetic fields bringing an evidence of minor extra superconducting phase with identical $T_c$ but different $H_{c2}$. If the phase with larger upper critical magnetic field is taken into account, the positive curvature of $\mu_0H_{c2}$ temperature dependence close to $T_c$ diminishes leading to good agreement of $\mu_0H_{c2}$$(T)$ with the WHH model.
The presence of multiple superconducting  phases, this time with different $T_c$s, is clearly visible on the surface of the studied sample, where our local STM/STS measurements reveal broad distribution of the superconducting energy gaps, which scale with $T_c$ with the same cooupling ratio as obtained from the heat capacity measurements.  We can conclude that there is only single intrinsic superconducting energy gap in Mo$_8$Ga$_{41}$ while other superconducting gaps  belong to stoichiometrically close phases.

\acknowledgements

This work was supported by the EU ERDF (European regional development fund) grant No. ITMS26220120047, by the Slovak Research and Development Agency, under Grant No. APVV-16-0372, by Slovak Scientific Grant Agency under contract VEGA-0149/16 and VEGA-0743/19, and by the U.S. Steel Ko\v sice, s.r.o. The work in Moscow was supported by the Russian Science Foundation, Grant No. 17-13-01033. V.Yu.V appreciates the support from the Mobilitas Program of the European Science Foundation, Grant No. MOBJD449.

\end{document}